\documentclass[11pt]{article}
\usepackage[a4paper, total={6in, 8in}]{geometry}
\usepackage{setspace}
\usepackage[T1]{fontenc}
\usepackage[utf8]{inputenc}
\usepackage{lmodern}

\usepackage[english]{babel}
\usepackage{csquotes}
\usepackage{graphicx}

\usepackage{natbib}

\begin{document}

	\title{Science in the Era of ChatGPT, Large Language Models and Generative AI: Challenges for Research Ethics and How to Respond}
	\date{}
	\author{Evangelos Pournaras\\
	School of Computing, University of Leeds, 
Leeds, UK}
	\maketitle

	\begin{abstract}
		
	Large language models of artificial intelligence (AI), such as ChatGPT, find remarkable but controversial applicability in science and research. This paper reviews epistemological challenges, ethical and integrity risks in science conduct in the advent of generative AI. This is with the aim to lay new timely foundations for a high-quality research ethics review. The role of AI language models as a research instrument and subject is scrutinized along with ethical implications for scientists, participants and reviewers. New emerging practices for research ethics review are discussed, concluding with ten recommendations that shape a response for a more responsible research conduct in the era of AI. 
		
	\end{abstract}
	
%	\doublespacing

	\section{Introduction}\label{sec:intro}
	
	Since the release of popular large language models (LLMs) such as ChatGPT, the transformative impact of artificial intelligence (AI) on broader society has been unprecedented. This is particularly alarming for science and its conquest of truth~\citep{Chomsky2023}. Generative AI and, particularly, conversational AI based on language models has set new ethical dilemmas for knowledge, epistemology and research practice. From authorship, to misinformation, biases, fairness and safety of interactions with human subjects, research ethics boards need to adapt to this new era in order to protect research integrity and set high-quality ethical standards for research conduct~\citep{Dis2023}. This paper focuses on reviewing these challenges with the aim of laying foundations for a timely and effective response. 
	
	ChatGPT is an AI chatbot released in November 2022 by OpenAI. It is a Generative Pre-trained Transformer (GPT), a type of artificial deep neural network with a number of parameters in the order of billions. It is designed to process sequential input data, i.e. natural language, without labeling (self-supervised learning), but with remarkable capabilities for parallelization that significantly reduce training time. The model is further enhanced by a combination of supervised and reinforcement learning based on past conversations as well as human feedback to fine-tune the model and its responses~\citep{Stiennon2020,Gao2022}. Other corporations follow with similar chatbots such as the one of Bard by Google. Generative AI expands beyond text, for instance, to images, videos and code~\citep{Cao2023}. 
	
	ChatGPT demonstrates powerful and versatile capabilities that are relevant for science and research. From writing and debugging software code to writing, translating and summarizing text, the quality of its output becomes indistinguishable from that of a human~\citep{Else2023}, while generating complex responses to prompts in a few seconds. Despite this success, AI language models suffer from hallucinations, an effect of producing plausible-sounding responses, which are nevertheless incorrect, inaccurate or even nonsensical. Illustratively, generative AI fails to abide by Asimov's three laws of robotics~\citep{Smith2023}: (i) Harmful outputs do occur (first law)~\citep{Wei2023,Davis2023}. (ii) Jailbroken prompts often result in both disobedience and harm (second law)~\citep{Wei2023}. (iii) New capabilities for autonomy, e.g. Auto-GPT~\citep{Yang2023}, and pervasiveness (integration on personal mobile devices) may create additional loopholes for conflicts to the first and second law (third law). 
	
	Disclaimers of ChatGPT state the following: "\emph{May occasionally generate incorrect information}", "\emph{May occasionally produce harmful instructions or biased content}", "\emph{Our goal is to get external feedback in order to improve our systems and make them safer}", "\emph{While we have safeguards in place, the system may occasionally generate incorrect or misleading information and produce offensive or biased content. It is not intended to give advice}", "\emph{Conversations may be reviewed by our AI trainers to improve our systems.}", "\emph{Please don't share any sensitive information in your conversations}" and "\emph{Limited knowledge of world and events after 2021}". 
	
	Each of these disclaimers reveal alerting implications of using AI language models in science. They oppose core values to support research integrity such as the concordat~\citep{Concordat2020} of the UK Research Integrity Office (UKRIO): (i) \emph{honesty in all aspects of research}, (ii) \emph{rigor in line with disciplinary standards and norms}, (iii) \emph{transparency and open communication}, (iv) \emph{care and respect for all participants, subjects, users and beneficiaries of research} and (v) \emph{accountability to create positive research environments and take action if standards fall short}.  Generative AI also challenges several of the Asilomar AI Principles~\citeyearpar{Asilomar2017}. 
		
	\cite{Chomsky2023} question the morality of asking amoral conversational AI moral questions, while~\cite{Awad2018} show empirical evidence about the cross-cultural ethical variations and deep cultural traits of social expectation from moral decisions of machines, i.e. the moral machine experiment. Generative AI runs the risks of copyright infringement and deskilling of early career researchers in scientific writing and research conduct~\citep{Gottlieb2023,Dwivedi2023}. Security threats in online experimentation can `pollute' human subject pools by replacing human subjects with conversational AI chatbots to claim compensations~\citep{Jansen2023,Wei2023}. Without safeguards for such new sources of misinformation, data quality and research conduct can be degraded at scale.
	
	AI language models also set foundational epistemological challenges addressing Karl Popper's seminal work on philosophy of science~\citep{Popper1959a,Popper1959b}. Can AI language models assist us to make scientific statements that are falsifiable, or are they rather preventing us from doing so within their opaque nature? Are we addressing reality by relying our scientific inquiry on them, and which reality is this? Do over-optimized AI language models subject to Goodhart's law~\citep{Manheim2018} manifest irrefutable truth? And if so, do these models constitute the wrong view of science that betrays itself in the craving to be right? 
	
	This paper dissects these questions with a focus on the research ethics review, although the discussion finds relevance to other facets of science such as education. To dissect the implications on science, the role of AI language models is distinguished as a \emph{research instrument} and \emph{research subject} when addressing a reseach hypothesis or questions related or not to generative AI. Moreover, the ethical challenges of AI digital assistance to \emph{scientists}, \emph{human research subjects} and \emph{reviewers} of research ethics are assessed. This scrutiny yields ten recommendations of actions to preserve and set new quality standards for research ethics and integrity as a response to the advent of generative AI.
	
	This paper is organized as follows: Section~\ref{sec:role-of-AI} reviews the different roles of generative AI in research design. Section~\ref{sec:digital-assistance} reviews the digital assistance provided by generative AI to scientists, participants and reviewers. Section~\ref{sec:application} discusses emerging research ethics review practices in the era of generative AI. Section~\ref{sec:recommendations} introduces ten recommendations to respond to the challenges of research ethics review. Finally, Section~\ref{sec:conclusion} concludes this paper and outlines future work.

	\section{The Role of Generative AI in Research Design}\label{sec:role-of-AI}

\begin{figure}[!htb]
		\centering
		\includegraphics[width=1.0\textwidth]{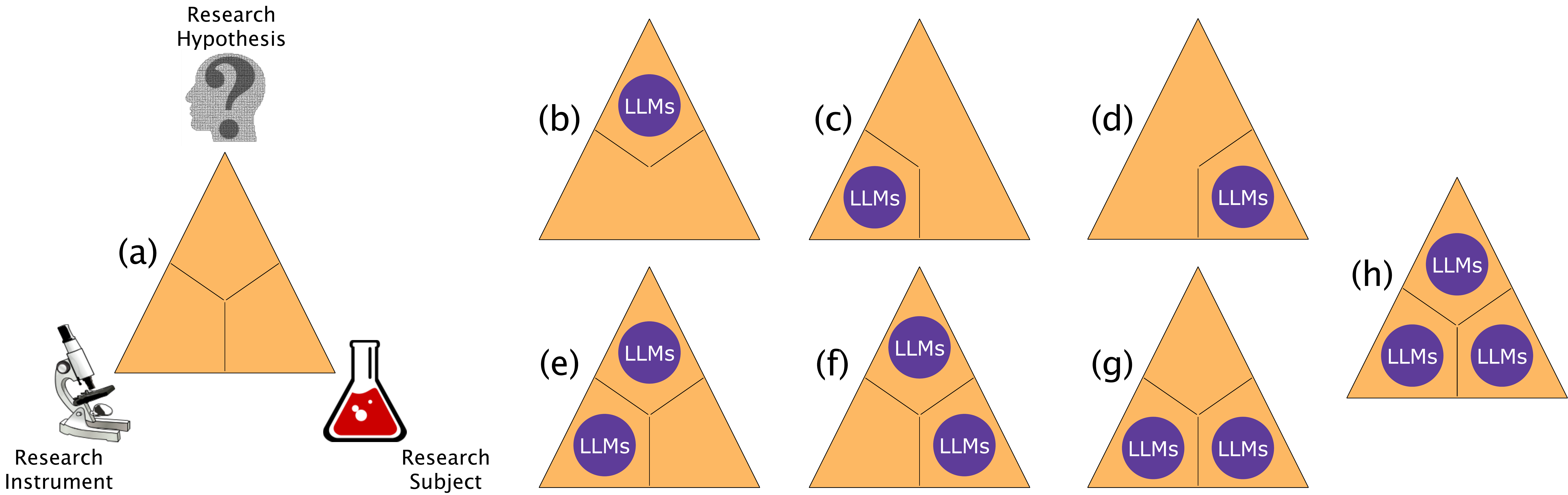}
		\caption{Generative AI such as large language models (LLMs) can be present in multiple stages of a research design within a research ethics application. Here we depict all combinations: (a) No generative AI models are involved. (b) Generative AI models can be the motivation behind formulating a research hypothesis or question. (c) They can also be used as a research instrument to acquire knowledge. (d) They can also be the research subject itself, when interacting with human research subjects or when acting independently. (e)-(h) Generative AI models may be involved in multiple stages of the research design. In this case, it becomes imperative to distinguish their role at each phase to dissect research integrity and ethical dilemmas that may not be apparent anymore. Note that in (c), (d), and (g), where AI language models do not motivate a research hypothesis or question but they are involved as a research instrument or subject, research integrity and ethical risks are likely to arise.}\label{fig:model}
	\end{figure}

	Within a research design serving a research hypothesis or question, generative AI can be involved as a research instrument or as a research subject, along with human subjects. This section distinguishes and discusses challenges and risks that may arise in these different contexts of a research ethics application. Figure~\ref{fig:model} illustrates where generative AI such as large language models can emerge in a research design. 
	
	\subsection{Generative AI as a research instrument}\label{sec:research-instrument}
	
	ChatGPT is documented as an emerging research instrument capable of writing manuscripts for publication, often controversially featured as a co-author~\citep{Connor2022,Transformer2022,Thorp2023,Else2023}, writing software code~\citep{Dwivedi2023} and collecting data via queries~\citep{Dwivedi2023}. Such tools are expected to come with capabilities for hypotheses generation in the future, including the design of experiments~\citep{Dis2023,Dwivedi2023}. Each of these instrumentations comes with different opportunities and challenges, including ethical ones. 
	
	During the design stage of research, including research ethics applications, there may be minimal support of AI language models on writing. However, the motivation of research, including literature review~\citep{Burger2023}, generation of hypotheses, research questions as well as identifying ethical dilemmas may be a result of interactions with conversational AI. Using the large capacity of conversational AI for knowledge summarization, these interactions can be systematized based on the Socratic method to foster intuition, creativity, imagination and potential novelty~\citep{Chang2023}. 
	
	However, often, creativity cannot be balanced with constraint~\citep{Chomsky2023}. At this stage, interactions with conversational AI require caution, running the risk of emulating or reinforcing a synergetic Dunning–Kruger effect~\citep{Gregorcic2023}: conversational AI may rely on limited (or wrong) knowledge, which, while presented as plausible to humans with similar limited knowledge, it may induce confirmation biases and diminish critical thinking. The mutual limitations of knowledge can be significantly underestimated in this context. 
	
	While research design choices may emerge from such interactions with conversational AI, a factual justification, a rigorous auditing process and moral judgments of these choices remain entirely under human premises (Recommendation 1 and 8 in Section~\ref{sec:recommendations}). Finding reliable sources, revealing data sources, accurate contextualization of facts and moral framing are not attainable at this moment, as they require both cognitive capabilities, accountability and transparency that current AI language models lack of (Recommendation 1, Section~\ref{sec:recommendations}). Whether existing ethics review processes are able to distinguish the risk level of research designs produced with the support of conversational AI as well as the mitigation actions, is an open question (Recommendation 5 in Section~\ref{sec:recommendations}). 
	
	During research conduct, integrity and ethical dilemmas may arise when using the direct output of conversational AI (knowledge acquisition) to confirm or refute a hypothesis, especially when this hypothesis is not about the AI system itself (see Figure~\ref{fig:model}c,~\ref{fig:model}d,~\ref{fig:model}g and Recommendation 4 in Section~\ref{sec:recommendations}). This output is in principle unreliable as it may contain incorrect or inaccurate information~\citep{Davis2023}. For instance, correct referencing may approach just 6\%~\citep{Blanco2022}. Moreover, AI language models tend to produce plausible content rather than content to be assessed as falsifiable, raising epistemological challenges~\citep{Popper1959a,Popper1959b}. The reliability of AI language models as effective proxies for specific human populations is subject of ongoing research~\citep{Argyle2023}. 
	
	Even if the output of AI language models is correct and accurate, it may not explain how such output is generated. For instance, there is often uncertainty to distinguish between lack of relevant data in the training set and failure to distill these data to credible information~\citep{Dis2023}. These models are usually black boxes with very low capacity to explain or interpret them. So far, this explainability is hard to assess for systems such as ChatGPT and Bard, which are closed and intransparent. This scenario may resemble an instrument collecting data exposed though to an unknown source of noise. Using instruments that have not passed quality assurance criteria may introduce various risks for users and work performed with such instruments and it is not different for AI language models. Standardized quality metrics are likely to arise for reporting to future research ethics applications (Recommendation 6 in Section~\ref{sec:recommendations}), for instance, the `algorithmic fidelity' that measures how well a language model can emulate response distributions from a wide spectrum of human groups~\citep{Argyle2023}.

	\subsection{Generative AI as a research subject}\label{sec:research-subject}
	
	The actual release of ChatGPT can be seen itself as a subject of research conducted by OpenAI with the aim to acquire user feedback that will improve AI language models. The initial interest lies on their actual capabilities to generate text and meaningful responses to user prompts. It also includes a discourse around their capabilities to perform calculations, write working code and jailbreaking via prompts that bypass the filters of its responses~\citep{Wei2023}. 
	
	While these initial investigations are mainly experimental and anecdotal, a rise of empirical research on ChatGPT is ongoing~\citep{Dwivedi2023,Kim2023,Bisbee2023}, e.g., survey research. However, this outbreak of empirical research is to a certain extent a byproduct of releasing a closed AI blackbox, with low capacity for explainability especially when the broader public does not have access to the model itself or the exact data with which it is trained. 
	
	OpenAI and other corporations may benefit from such research as a (free) crowd-sourcing feedback to calibrate their products, without sharing responsibility for doing so. Nonetheless, this may not be the original aims and intentions of scientists conducting such research. Such misalignment comes with ethical considerations on the value of this research and requires a critical stand by researchers and research ethics reviewers (Recommendation 7 in Section~\ref{sec:recommendations}). While the methods of research on human subjects are well established (e.g. statistical methods, sociology, psychology, clinical research), the methods on AI subjects remain of different nature, pertinent to engineering and computer science. As human and AI subjects become more interactive, pervasive, integrated and indistinguishable, research ethics reviews need to account for (and expect) inter-disciplinary mixed-mode research methods (Recommendation 2 in Section~\ref{sec:recommendations}).

	\section{Digital Assistance by Generative AI}\label{sec:digital-assistance}

	AI language models can provide assistance to scientists, participants in human experimentation as well as to reviewers of research ethics applications. This section assesses ethical challenges pertinent to these beneficiaries. 
	
	\subsection{AI-assisted scientist}\label{sec:scientist}
	
	As introduced in Section~\ref{sec:role-of-AI}, the support of AI language models to scientists for literature review, writing papers, code, collecting data and performing experiments involves several challenges of integrity and ethics/moral. One question that may arise is how generative AI can contribute to the making of future scientists. Can they be part of the education of PhD students or will they result in deskilling, especially when students are not familiar with academic norms~\citep{Dwivedi2023}? Will such models be able to provide any level of self-supervision capability? The feasibility of research designs, success prediction of research proposals and reviewing manuscripts at early stages and before submission to journals, are some examples in which linguistics, epistemology and theory of knowledge set limits that for AI language models is hard to overcome~\citep{Chomsky2023}.

	\subsection{AI-assisted participant}\label{sec:participant}	
	
	Studying human research subjects assisted by AI language models requires a highly interdisciplinary perspective to dissect the ethical challenges and risks that may be involved (Recommendation 2 in Section~\ref{sec:recommendations}). Such studies may aim to address the human subjects (i.e. social science), the AI language models when interacting with humans (i.e. computer science, decision-support systems), or both (e.g. human-machine intelligence). Design choices in AI systems for digital assistance to humans have direct ethical implications. 
	
	For instance, access to personal data for training AI models, centralized processing of large-scale sensitive information by untrustworthy parties and intransparent algorithms that reinforce biases, discrimination and informational filter bubbles pose significant risks. These include loss of personal freedoms and autonomy by manipulative algorithmic nudging, which participants may experience directly under research conduct, as well as broader implications in society~\citep{Hine2021} related to environment, health and democracy~\citep{Pournaras2023,Asikis2021,Helbing2021,Helbing2023}. The use of emerging open language models provides higher transparency to address some of these challenges~\citep{Patel2023,Scao2022}. Privacy-preserving interactions with AI language models, comparable to browsing with the DuckDuckGo search engine, are required (Recommendation 3 in Section~\ref{sec:recommendations}).
	%	https://bigscience.huggingface.co RECOMMENDATIONS? 
	
	Participants need to be informed about these risks when participating in such studies. For instance, information consent needs to account for any sensitive information shared during interactions with ChatGPT. Researchers do not have full control of the data collected in the background by OpenAI. As a result, participants need to be informed about the terms of use of AI language models. Moreover, responses by AI language models require moderation by researchers if they are likely to cause any harm to participants or special groups. Research ethics applications need to reflect and mitigate such cases (Recommendation 9 in Section~\ref{sec:recommendations}). 

	\subsection{AI-assisted reviewer}\label{sec:reviewer}
	
	The support of generative AI to research ethics reviewers is a highly complex matter that perplexes both ethical matters within research communities as well as moral matters of individual reviewers. People do not share the same judgments between the ethical choices of a human or a machine~\citep{Hidalgo2021}.	
	
	AI language models show limited capabilities for ethical positioning, let alone moral positioning, possessing an apathy and indifference to implications of ethical choices~\citep{Chomsky2023}. They can endorse both ethical and unethical choices based on correct and incorrect information~\citep{Chomsky2023}. Nevertheless, they manage to influence users' moral judgments in an non-transparent way~\citep{Krugel2023}. 
	
	On the other hand, AI models can be used to effectively detect plagiarism or to perform pattern matching tasks that do not involve complex explanations or analysis of consequences. For instance, GPTZero is able to distinguish between text generated by humans vs. AI language models~\citep{Heumann2023}, which would be otherwise hard for reviewers to distinguish~\citep{Else2023}. Moreover, AI language models can assist reviewers, whose research background may be in a different discipline than the one of the proposed research. Summarizing necessary background knowledge and providing summaries in layman's terms can benefit research ethics reviewers~\citep{Hine2021} as long as they remain critical on the generated output of AI language models. 
		
	As a result, AI language models are far from replacing reviewers in distilling ethical and moral implications of a research design, nevertheless, they can still play a role in the reviewing process by automating processes for pattern matching or making necessary background knowledge more accessible to reviewers, who may lack thereof.

	\section{Research Ethics Review Practices}\label{sec:application}
	
	The need for regulatory and procedural reforms in research ethics review as a response to challenges of Big Data and data subjects dates back before generative AI~\citep{Ferretti2021,Hine2021}. Currently, the scope and practices of research ethics review are becoming broader and more multifaceted to cover the new alarming risks of generative AI. Two factors distinguish these research ethics review practices: (i) \emph{scale of impact} and (ii) \emph{stage of research}.  
	
	Institutional review boards for research ethics mainly address the impact of generative AI on human participants before the research conduct. Broader implications of the research on society are not explicitly addressed, although initial results from piloting an \emph{Ethics and Society Review}~\citep{Bernstein2021} as a requirement to access funding show a positive impact~\citep{Bernstein2021}. During research conduct, research ethics reviews mainly address any required adjustments in the research design rather than other unanticipated risks emerging from the application or new developments of AI. 
	
	Moreover, new research ethics review practices have been established for funding institutions~\citep{Bernstein2021}, conferences and journals~\citep{Srikumar2022}. These include (i) \emph{impact statements}, (ii) \emph{checklists} and (iii) \emph{code of ethics or guidelines}. Impact statements include ethical aspects, questions and future positive or negative societal consequences, as well as identification of human groups, behavioral and socio-economic data. Checklists are used to flag papers for additional ethics review by an appointed committee, while code of ethics and guidelines support reviewers to flag papers that violate them. 
	
	While there is evidence that such practices can support panels to identify risks related to the harming of subgroups and low diversity~\citep{Bernstein2021}, encouraging research communities to apply universal practices in different contexts and disciplines is a highly complex endeavor, given the current rapid AI developments and the unanticipated impact of these on society (Recommendation 10 in Section~\ref{sec:recommendations}). 
	
	There are particular aspects of existing research ethics applications dealing with human aspects that are perplexed with the use of generative AI. These include individuals who can or cannot consent to terms of use and conditions of generative AI software, participants with disabilities, vulnerable groups and children, exclusion of certain groups, deception and incomplete disclosure, short and long term risks of participation, protection of personal data, anonymity and data storage. Research ethics review needs to address explicitly any additional risks involved in those aspects by using generative AI.

	\section{Ten Recommendations for Research Ethics Committees}\label{sec:recommendations}
	
	This section introduces ten recommendations for research ethics committees. They distill the challenges and responses to AI language models involved in research ethics applications. They significantly expand on other earlier recommendations~\citep{Hine2021} such as the one of World Association of Medical Editors (WAME) mainly addressing authorship, transparency and responsibility~\citep{Zielinski2023}. They also constitute actions within the broader recommendations made for (i) studying community behavior and share learnings, (ii) expanding experimentation of ethical review and (iii) creating venues for debate, alignment and collective action~\citep{Srikumar2022}. The ten recommendations are summarized as follows: 
	
	\begin{enumerate}
		\item Humans should always remain accountable for every scientific practice. %Why this is particulrly important here comapred to SPSS? Example on AI explainability ! 
		\item An interdisciplinary panel of reviewers should be employed to assess research ethics applications with elements on generative AI. 
		\item The use of generative AI models, their version, prompts and responses need to be documented and reported in any phase of the planned research. As a response, ethics reviews should detect potential inaccuracies, biases and inappropriate referencing. Mitigation by encouraging and promoting open generative models can improve accountability and transparency. 
		\item Research ethics applications that aim to address research hypotheses and questions out of the scope of generative AI, which do involve generative AI models as a research instrument or subject, are likely to involve research integrity and ethics issues and should be treated as high-risk applications. 
		\item Ethics review applications require new criteria and practices to distinguish low and high integrity risks in research designs produced with the support of generative AI. Determining appropriate mitigation actions to account for different risk levels is required. 
		\item Researchers who engage with generative AI in their research should report their countermeasures against inaccuracies, biases and plagiarism. Ethical review applications need to cover these risks.
		\item The motivation and aim of research on generative AI should come with merit and go beyond testing of prompts lacking a rigorous scientific inquiry. 
		\item Auditing protocols are required for each input to generative AI models that are closed and proprietary, as a way to prevent sharing sensitive personal or proprietary information of researchers or participants. 
		\item Any output of generative AI that may harm participants or is sensitive to special groups requires moderation by researchers. Informed consent to relevant terms of use of generative AI models is required. 
		\item Communities on research ethics and regulatory bodies require to maintain an agreement on AI language models that can be used or should not be used in research. For instance, models that are obsolete, inaccurate, highly biased and violate values of science conduct shall be excluded, replaced or used with significant caution. 
	\end{enumerate}
	
	These recommendations should be used as an open and evolving agenda rather than a final list of actions. The current landscape of AI language models and research ethics remains multi-faceted, rapidly changing and complex. Timely adjustments are needed as a response.

	\section{Conclusion and Future Work}\label{sec:conclusion}
	
	To conclude, the challenges and risks of generative AI models for science conduct are highly multifaceted and complex. They are not yet fully understood, as developments are fast with significant impact and unknown implications. 
	
	Research ethics boards have a moral duty to follow these developments, co-design necessary safeguards and provide a research ethics review that minimizes ethical risks. A deep interdisciplinary understanding of the role that AI language models can play in all stages of research conduct is imperative. This can dissect ethical challenges involved in the digital assistance of scientists, research participants and reviewers. 
	
	The ten recommendations introduced in this paper set an agenda for a dialogue and actions for more responsible science in the era of AI.

	\section*{Acknowledgments}\label{sec:ack}
	
	Thanks to Maria Tsimpiri for inspiring discussions. Evangelos Pournaras is supported by a UKRI Future Leaders Fellowship (MR\-/W009560\-/1): `\emph{Digitally Assisted Collective Governance of Smart City Commons--ARTIO}', an Alan Turing Fellowship and the SNF NRP77 `Digital Transformation' project "Digital Democracy: Innovations in Decision-making Processes", \#40\-77\-40\_\-18\-72\-49, the SNF NRP77 project `Digital Transformation' project "Digital Democracy: Innovations in Decision-making Processes", \#407740\_187249 as well as the European Union, under the Grant Agreement GA101081953 attributed to the project H2OforAll—\emph{Innovative Integrated Tools and Technologies to Protect and Treat Drinking Water from Disinfection Byproducts (DBPs)} . Views and opinions expressed are, however, those of the author(s) only and do not necessarily reflect those of the European Union. Neither the European Union nor the granting authority can be held responsible for them. Funding for the work carried out by UK beneficiaries has been provided by UK Research and Innovation (UKRI) under the UK government’s Horizon Europe funding guarantee [grant number 10043071].

	\bibliography{paper}
	\bibliographystyle{apsr}

	\end{document}